\documentclass[conference]{IEEEtran}
\IEEEoverridecommandlockouts
\usepackage{amsmath,amssymb,amsfonts}
\usepackage{listings}
\usepackage{algorithm}
\usepackage{algorithmicx}
\usepackage[noend]{algpseudocode}
\usepackage{graphicx}
\usepackage{textcomp}
\usepackage{xcolor}
\usepackage{tikz}
\usetikzlibrary{shapes,positioning,calc}
\usepackage[style=ieee, bibencoding=utf8, citestyle=numeric-comp]{biblatex}
\usepackage[unicode, hidelinks]{hyperref}
\def\BibTeX{{\rm B\kern-.05em{\sc i\kern-.025em b}\kern-.08em
    T\kern-.1667em\lower.7ex\hbox{E}\kern-.125emX}}
\usepackage{booktabs}
\usepackage{colortbl}
\usepackage{multirow}
\usepackage{flushend}
\begin{document}

\title{Sydr: Cutting Edge Dynamic Symbolic Execution
\thanks{This work was supported by RFBR grant 20-07-00921 A.}}

\author{
\IEEEauthorblockN{
  Alexey Vishnyakov\IEEEauthorrefmark{1},
  Andrey Fedotov\IEEEauthorrefmark{1},
  Daniil Kuts\IEEEauthorrefmark{1},
  Alexander Novikov\IEEEauthorrefmark{1},
  Darya Parygina\IEEEauthorrefmark{1}\IEEEauthorrefmark{2}, \\
  Eli Kobrin\IEEEauthorrefmark{1}\IEEEauthorrefmark{2},
  Vlada Logunova\IEEEauthorrefmark{1}\IEEEauthorrefmark{3},
  Pavel Belecky\IEEEauthorrefmark{1} and
  Shamil Kurmangaleev\IEEEauthorrefmark{1}
}
\IEEEauthorblockA{
  \IEEEauthorrefmark{1}Ivannikov Institute for System Programming of the RAS
}
\IEEEauthorblockA{
  \IEEEauthorrefmark{2}Lomonosov Moscow State University
}
\IEEEauthorblockA{
  \IEEEauthorrefmark{3}Moscow Institute of Physics and Technology
}
Moscow, Russia \\
\{vishnya, fedotoff, kutz, a.novikov, pa\_darochek, kobrineli, vlada, belecky, kursh\}@ispras.ru
}

\maketitle

\begin{tikzpicture}[remember picture, overlay]
\node at ($(current page.south) + (0,0.65in)$) {
\begin{minipage}{\textwidth} \footnotesize
  Vishnyakov A., Fedotov A., Kuts D., Novikov A., Parygina D., Kobrin E.,
  Logunova V., Belecky P., Kurmangaleev Sh. Sydr: Cutting Edge Dynamic Symbolic
  Execution. 2020 Ivannikov ISPRAS Open Conference (ISPRAS), IEEE, 2020, pp.
  46-54. DOI: \href{https://www.doi.org/10.1109/ISPRAS51486.2020.00014}{10.1109/ISPRAS51486.2020.00014}.

  \copyright~2020 IEEE. Personal use of this material is permitted. Permission
  from IEEE must be obtained for all other uses, in any current or future media,
  including reprinting/republishing this material for advertising or promotional
  purposes, creating new collective works, for resale or redistribution to
  servers or lists, or reuse of any copyrighted component of this work in other
  works.
\end{minipage}
};
\end{tikzpicture}

\begin{abstract}
The security development lifecycle (SDL) is becoming an industry standard.
Dynamic symbolic execution (DSE) has enormous amount of applications in computer
security (fuzzing, vulnerability discovery, reverse-engineering, etc.). We
propose several performance and accuracy improvements for dynamic symbolic
execution. Skipping non-symbolic instructions allows to build a path predicate
1.2--3.5 times faster. Symbolic engine simplifies formulas during symbolic
execution. Path predicate slicing eliminates irrelevant conjuncts from solver
queries. We handle each jump table (switch statement) as multiple branches and
describe the method for symbolic execution of multi-threaded programs. The
proposed solutions were implemented in Sydr tool. Sydr performs inversion of
branches in path predicate. Sydr combines DynamoRIO dynamic binary
instrumentation tool with Triton symbolic engine. We evaluated Sydr features on
64-bit Linux executables.
\end{abstract}


\section{Introduction}

We can search errors in programs in various ways: at the compile time, manually,
applying static analysis tools to source~\cite{bessey10, ivannikov14} and
binary~\cite{balakrishnan05, aslanyan20} code, dynamic analysis tools,
formal verification tools, etc. Security researchers and developers
widely use fuzzing~\cite{sargsyan19, fioraldi20}, dynamic
symbolic execution~\cite{chipounov12, shoshitaishvili16}, and systems combining both of
them~\cite{cha12, stephens16, yun18} to
detect errors. The security development lifecycle (SDL) is becoming an industry
standard~\cite{howard06, iso08, gost16}. Developers are required to apply various analysis tools to
improve the quality of their product. These tools have two purposes:
(1)~generate new inputs that expand the code coverage; (2)~find errors. Of
course, while solving (1)~problem one can detect a certain number of errors, but
it is more efficient to separate these tasks. Based on the experience of
developing dynamic symbolic execution tools~\cite{isaev10, gerasimov17,
padaryan15}, we developed a new dynamic symbolic execution tool (Sydr) that
addresses the problem of expanding coverage. We are going to extend Sydr to
solve problem of finding errors in future.

Dynamic symbolic execution~\cite{king76, schwartz10, baldoni18} explore variation of
initial input data on some fixed execution path. Initially each byte of input
data is modeled by a free symbolic variable. Each instruction is modeled with
an SMT~\cite{smt-lib} formula over constants and symbolic variables according to
its operational semantics. Symbolic engine maintains a symbolic state that is a
mapping from memory bytes and registers to SMT formulas. All symbolic
register/memory modifications update the symbolic state with new formulas.
Branch conditions on the explored path are represented by SMT formulas and form
the path predicate. Thus, the path predicate contains the constraints that
describe the explored path. The solution to conjunction of these path constraints
is an input data that follows the same execution path. In order to invert some
branch, we negate its path constraint.

We symbolically execute a program to invert branches. Thus, we are able to
discover new paths that regular fuzzing can hardly reach. The main goal of this
work is improving dynamic symbolic execution accuracy and performance. Accuracy
is essential, because we need generated inputs to actually invert target
branches and discover new paths. Increasing performance helps to gain new inputs
faster.

This paper makes the following contributions:
\begin{itemize}
  \item We surveyed existing symbolic execution improvement approaches and
  implemented the most promising ones. We evaluated each method to measure
  its impact on symbolic execution in general.
  Sections~\ref{Skip_nonsymbolic}--\ref{Multithreading} describe the main
  features: skipping non-symbolic instructions~\ref{Skip_nonsymbolic}, AST
  simplification~\ref{AST_simplification}, path predicate slicing~\ref{Slicing},
  indirect jumps resolving~\ref{Jump_tables}, and handling
  multi-threaded programs~\ref{Multithreading}. The experimental results
  presented in Section~\ref{Evaluation}.
  \item We present Sydr, a dynamic symbolic execution tool, which implements
  each of these techniques and can be used as a path explorer for fuzzing
  tools. We describe the tool design in Section~\ref{Design}.
\end{itemize}

\section{Skipping Non-Symbolic Instructions}

\label{Skip_nonsymbolic}

Symbolic execution approximately slows down a target application execution
between 1000 and 1000000 times. We skip symbolic execution of non-symbolic
instructions to build the path predicate faster and reduce memory usage. To
determine whether an instruction is symbolic we retrieve all its explicit and
implicit (e.g. \texttt{pop rax} implicitly reads from the stack and modifies the
stack pointer) operands with the help of DynamoRIO~\cite{bruening04, dynamorio}
disassembler. Then we symbolically execute the instruction iff any of its
read/write registers (including ones used for computing memory address, e.g.
\texttt{rbx} in \texttt{mov rax, [rbx]}), memory, or flags are symbolic.
Skipping non-symbolic instructions allows us to build path predicate 1.2--3.5
times faster (Table~\ref{tbl:path-predicate}). Moreover, we consume less memory
because we create less SMT~\cite{smt-lib} statements.

\section{AST Simplification}

\label{AST_simplification}

Symbolic engines tend to use an intermediate AST representation that is later
translated to SMT. Symbolic engine may simplify these ASTs before passing them
to solver~\cite{shoshitaishvili16, claripy}. We implemented several AST
simplifications in Triton~\cite{saudel15, triton} symbolic engine. These
simplifications improve accuracy of symbolic execution, help solver, reduce
memory used by ASTs, and improve readability of printed SMT statements. We list
some of them below:
\begin{itemize}
  \item $A \mathbin{\&} A \rightarrow A$, $A \mathbin{|} A \rightarrow A$.
  \item $A \oplus A \rightarrow 0$, $0 * A \rightarrow 0$,
    $A - A \rightarrow 0$, $0 \mathbin{\&} A \rightarrow 0$,
    $0 \ll A \rightarrow 0$, $0 \gg A \rightarrow 0$. Triton marks AST as
    symbolic iff it contains symbolic variables. AST simplifications can remove
    symbolic marks after kill operations. For instance, symbolic \texttt{rax}
    after \texttt{xor rax, rax} will be marked as non-symbolic.
  \item \begin{small}
        \texttt{((\_~extract~high~low) ((\_~extract~hi~lo) A)) $\rightarrow$
                ((\_~extract~high+lo~low+lo)~A)}
        \end{small}
  \item \begin{small}
        \texttt{((\_~extract~11~9) (concat (\_~bv1~8) (\_~bv2~8) (\_~bv3~8) (\_~bv4~8))) $\rightarrow$
                ((\_~extract~3~1) (\_~bv3~8))}\end{small}.
    Triton represents each symbolic parent register with an AST. Modification of
    lower register byte results in AST that concatenates extracted register high
    part with a new byte value. In particular, if we place a constant in lower
    byte of symbolic register and later extract it, we will get a symbolic AST.
    The proposed simplification provides a non-symbolic AST. So, \texttt{jz}
    branch in instruction sequence \texttt{mov rax, symbolic\_variable ; mov al,
    0x00 ; test al, al ; jz 0xdeadbeef} won't be symbolic.
  \item \begin{small}
        \texttt{(concat ((\_~extract~31~24)~A) ((\_~extract~23~16)~A)
                        ((\_~extract~15~8)~A) ((\_~extract~7~0)~A)) $\rightarrow$
                ((\_~extract~31~0)~A)}\end{small}.
    For instance, this simplification is useful when we store register in memory
    and later retrieve it. Triton extracts each register byte and stores it to
    symbolic memory. During the register load Triton concatenates all its
    extracted parts.
  \item \begin{small}
        \texttt{((\_~extract~31~0) ((\_~zero\_extend~32) (\_~bv1~32))) $\rightarrow$ (\_~bv1~32)}\end{small}.
    Instructions that operate on 32-bit general-purpose registers in x86-64 zero
    extend the result to parent register. If we move one 32-bit register to
    another and ask for its AST, we are going to get lower 32 bits extraction
    from zero extended register. The simplification just returns the original
    moved register.
\end{itemize}

\section{Path Predicate Slicing}

\label{Slicing}

\begin{algorithm}[t]
  \caption{Path predicate slicing algorithm.}
  \textbf{Input:} $cond$~-- predicate for target branch inversion,
  $\Pi$~-- path predicate (path constraints prior to the target branch).
  \begin{algorithmic}
    \State $vars \gets used\_variables(cond)$
    \Comment{slicing variables}
    \State $change \gets vars$
    \While {$change \neq \varnothing$}
      \State $change \gets vars$
      \ForAll {$c \in \Pi$} \Comment{iterate over path constraints}
        \If {$vars \cap used\_variables(c) \neq \varnothing$}
          \State $vars \gets vars \cup used\_variables(c)$
        \EndIf
      \EndFor
      \State $change \gets vars \setminus change$
    \EndWhile
    \State $\Pi_S \gets cond$
    \Comment{predicate for branch inversion}
    \ForAll {$c \in \Pi$} \Comment{iterate over path constraints}
    \If {$vars \cap used\_variables(c) \neq \varnothing$}
      \State $\Pi_S \gets \Pi_S \wedge c$
    \EndIf
    \EndFor
    \State \textbf{return} $\Pi_S$
  \end{algorithmic}
  \label{alg:slicing}
\end{algorithm}

We use path predicate slicing (a.k.a. constraint independence
optimization~\cite{cadar06, cadar08}) to eliminate irrelevant conjuncts from
solver queries. For clarity, we define a path predicate $\Pi$ as a sequence of
constraints. Each constraint corresponds to a taken branch condition on the
execution trace. Algorithm~\ref{alg:slicing} returns a predicate for inversion
of the target branch. $\Pi$ contains constraints for all taken branches that
were executed before the target branch. Function $used\_variables(c)$ returns a
bitset of symbolic variable numbers used in constraint $c$. We initialize
slicing variables $vars$ with variables used in predicate for target branch
inversion $cond$ (a negation of taken target branch constraint). Then we iterate
over and over constraints in $\Pi$ updating $vars$ with variables that
transitively depend on slicing variables. Finally, we conjunct only those branch
constraints that have variables from $vars$. Thus, we leave only constraints
that are relevant to inverting the target branch.

With slicing applied solver returns a model only for some subset of symbolic
variables. We retrieve values for missing symbolic variables from initial input
data. The resulting solution is correct due to the fact that initial input data
is already a solution for a path predicate.

Slicing allows to perform a more powerful symbolic execution. It has the
following advantages:
\begin{enumerate}
  \item Solver consumes less memory and time to resolve a query. We get model
    only for a part of input data that is responsible for branch inversion.
  \item Undertaint~\cite{kang11} can cause some symbolic variables to be
    underconstrained. Thus, generated input may not reproduce the desired path.
    Slicing remove possibly underconstrained symbolic variables from the solver
    query. These variables values are taken from the initial input.
\end{enumerate}

Consider the following code:
\begin{lstlisting}[language=C, basicstyle=\small\ttfamily, numbers=left,
                   xleftmargin=2em]
char* syms = "SLICING FIX IT!\n";

int main(int argc, char **argv) {
  FILE *ptr = fopen(argv[1], "rb");
  unsigned *b = malloc(6 * sizeof(int));
  fread(b, 6, sizeof(int), ptr);
  int len = strlen(syms);
  if (b[0] < len)
    if (syms[b[0] % len] == '!')
      if (b[2] > '@')
        if (b[5] + b[4] < 'B')
          if (b[3] + b[5] > '@')
            if (b[1] + b[3] > '@')
              if (b[4] < '9')
                if (b[1] > '@')
                  printf("OK\n");
                else
                  printf("FAIL\n");
}
\end{lstlisting}
Initial input data leads to printing \texttt{FAIL} in line 18. We illustrate how
slicing algorithm inverts the branch in line 15. Slicing appends constraints for
the following branches to the resulting predicate:
\begin{itemize}
  \item Line 15 (slicing variable \texttt{b[1]}).
  \item Line 13 (slicing variables \texttt{b[1]}, \texttt{b[3]}).
  \item Line 12 (slicing variables \texttt{b[1]}, \texttt{b[3]}, \texttt{b[5]}).
  \item Line 11 (slicing variables \texttt{b[1]}, \texttt{b[3]}, \texttt{b[4]},
    \texttt{b[5]}).
  \item Line 14 (slicing variables \texttt{b[1]}, \texttt{b[3]}, \texttt{b[4]},
    \texttt{b[5]}).
\end{itemize}
Line 9 contains a symbolic address \texttt{b[0] \% len} that will be concretized
by the symbolic engine. So, the branch in line 9 is underconstrained (not
symbolic). Appending branch 8 to predicate without 9 may result in invalid
\texttt{b[0]} model. Slicing skips branch 8. Thus, we successfully generate an
input causing our example program to print \texttt{OK}. If we run the example on
input generated without slicing, condition in line 9 won't hold.

\section{Indirect Control Transfers Resolving}

\label{Jump_tables}

Handling indirect control flow transfers is crucial for the complete and
accurate program analysis. For the branch inversion problem, we are only
interested in table control flow transitions. In such cases, the target jump
address is taken from an array of pointers located in the program memory. The
offset for the corresponding array element is computed from the branch
condition. Jump tables~\cite{cifuentes01} are generated by compilers from the
long switch and if-else statements for the optimization purpose. Furthermore,
compiler also produces jump tables for function pointer arrays. Besides the
direct code pointers, jump tables may contain values (address offsets) that take
part in computing the target jump address:

\begin{small}
\begin{verbatim}
lea rdx, [rax * 4]
lea rax, [rip + 0x155]
mov eax, [rdx + rax]
movsxd rdx, eax
lea rax, [rip + 0x148]
add rax, rdx
jmp rax
\end{verbatim}
\end{small}

The assembly above is a typical indirect jump. In the first two lines the table
index and the base address are calculated. In the next two lines a value loaded
from the jump table and stored in \texttt{rdx} register. Then this value is
added to the computed target base address and then a jump is made to the resulting
 address.

To dereference a symbolic pointer Triton~\cite{saudel15, triton} gets its value
from the concrete state. We propose indirect control transfer resolving to
partially handle symbolic pointers.

We consider each jump on address calculated from memory cell as a potential
indirect control transfer instead of determining the table control flow
dependencies by code patterns. We perform backward slicing~\cite{weiser84}
within a current basic block to detect such transitions. It's a trivial case
when jump/call instruction has memory reference operand. If target operand is a
register, we start tracking this register data flow dependencies up to the
beginning of the basic block. Thus, we locate an instruction reading value from
memory that forms the jump target register.

Firstly, we check whether jump table exists at the previously detected
instruction memory access address. We support two kinds of tables which
contain addresses or offset values. Address tables should contain values that
are valid executable addresses. We use heuristic for the offset tables. An
offset should be a negative double word value. At least one of
adjacent memory cells should contain value of the same type, otherwise
it will not be interpreted as a valid jump table. We assume that table is
continuously located in memory. We parse memory in both directions from the
current access while the conditions for corresponding table type are met.
Parsing stops upon reaching the configurable maximum table size limit. Besides,
the valid jump table should contain at least three entries.

Sometimes several different jump tables are placed in memory continuously. In
such cases the exact table bounds cannot be determined, and the maximum size
limitation prevents memory parsing overhead. As a result, part of the current
jump table is missed and part of adjacent jump table is parsed. Missed part of
jump table can be parsed during re-execution of analyzed program with another
memory access address. And those jump table entries, which belong to another
indirect jump will produce incompatible path constraints during branch
inversion.

After successful jump table parsing, we generate path constraints for the
indirect jump. A condition for each branch is an equality of the symbolic
pointer expression and the corresponding jump table entry address. Since the
indirect jumps are usually compiled from switch statements, some jump table
entries can point to the same jump target. We create only one path constraint
for each unique jump target to prevent generating several different inputs for
the one jump direction. The conditions for duplicated targets are merged with
disjunction.

If jump table contains offset values, we should calculate jump table
targets separately. Usually target addresses for such cases are computed as some
base address plus offset value from the table. Knowledge of concrete target address
for the current execution and offset value for the current branch helps to
determine this base address. Then destination addresses for every jump table
branch can be easily deduced by adding offset to this computed base.

\section{Symbolic Execution of Multi-Threaded Programs}

\label{Multithreading}

The regular analysis of multi-threaded programs corrupts symbolic model and makes
the following symbolic execution incorrect. To be able to analyze such programs,
we need to keep track of thread switching and maintain separate symbolic states
for each thread.

All threads within one process have all memory shared but their own register values,
so this should be considered in symbolic model. All threads can also have a
shared path predicates storage for symbolic branches, because they are built on different
symbolic registers values and won't affect each other.

The process of saving and restoring registers on the control flow transition
between threads is called a context switch. In order to handle multi-threaded
programs we implemented a context switching operation on symbolic model. We
maintain a thread contexts storage that contains symbolic registers for each
thread. On each thread switching we save all symbolic registers and replace them
with symbolic registers for the current thread.

The proposed technique application can be considered on the following program:
\begin{lstlisting}[language=C, basicstyle=\small\ttfamily, numbers=left,
xleftmargin=2em, breaklines=true]
int d[20], mins[4], P[4] = {0, 1, 2, 3};

void *min(void *thread_number) {
  int i = *((int *) thread_number);
  int cnt = sizeof(d) / sizeof(*d) / 4;
  mins[i] = d[i * cnt];
  for (int j = i*cnt+1; j < (i+1)*cnt; ++j)
    if (mins[i] > d[j]) mins[i] = d[j];
}

int main(int argc, char **argv) {
  int fd = open(argv[1], O_RDONLY);
  read(fd, d, sizeof(d));
  pthread_t t[4];
  for (int i = 0; i < 4; ++i)
    pthread_create(t+i,0,min,(void*)&P[i]);
  for (int i = 0; i < 4; ++i)
    pthread_join(t[i], 0);
  int m = mins[0];
  for (int i = 1; i < 4; ++i)
    if (m > mins[i]) m = mins[i];
  if (m > 100) printf("min>100");
}
\end{lstlisting}
This example implements a minimum search in an input array. In lines 15--16 four
threads are created. Each thread searches a minimum for a part of an input
array. Afterwards, the main thread computes a global minimum. We symbolically
execute this program on array with minimum less than 100. In lines 6--8 local
minimums are stored to the shared symbolic memory. In lines 19--21 the main
thread computes a global minimum. We are inverting the branch in the line 22.
Thus, the program prints \texttt{min>100}. Furthermore, the example illustrates
that dynamic symbolic execution is a path-sensitive analysis. The generated
input will not only have numbers greater than 100, but also it will satisfy all
constraints from branches in lines 8 and 21. These constraints actually define a
partial order on an input array. If we don't switch symbolic registers, we get
additional unsound path constraints and lose
some essential constraints on input array. It is due to the fact that symbolic registers
get overwritten by ones from the other thread. Thus, some array elements may be
less than 100 and the generated input won't invert the target branch.

The limitation of this approach is that we don't influence the thread order
during program execution. A generated input may not follow an expected path. To
solve this problem it is necessary to implement a thread scheduler to
arrange threads order~\cite{guo16}. We may address this problem in future.

\section{Implementation}

\label{Design}

We implemented the improvements described above in Sydr (Symbolic DynamoRIO) tool. Sydr is a dynamic
symbolic execution (DSE) tool based on dynamic binary instrumentation (DBI).
Sydr performs symbolic execution along one path (defined by input data) and
generates new inputs that invert branches discovered on that path.

There are two approaches for implementing DSE: (1)~collect execution trace and
perform symbolic execution using that trace~\cite{godefroid08, padaryan15};
(2)~perform symbolic execution while program is executed~\cite{molnar07}. The
(1)~method has an overhead for storing execution trace on hard drive and
processing the trace to generate SMT formulas. The technique~(2) doesn't have
overhead for storing traces on disk, but it is also challenging. DBI allows to
insert analyzing code before every executed instruction. This instrumentation
code may drive symbolic execution. Such analysis is limited to 4GB RAM when it
is applied to 32-bit executables. Moreover, problems may occur when your
instrumentation code is complex and it uses some external libraries. For
instance, DynamoRIO client crashes when it is linked with \texttt{pthread}
library~\cite{drpthread}. Furthermore, DynamoRIO heap is quite slow~\cite{drheap}. These
kind of problems motivated us to separate concrete and symbolic execution into
two processes. This separation allows to reduce concrete executor
code. Symbolic executor is not limited to 4~GB RAM and can be linked with any
libraries needed for analysis.

\begin{figure}[t]
    \centering
    \includegraphics[width=\linewidth]{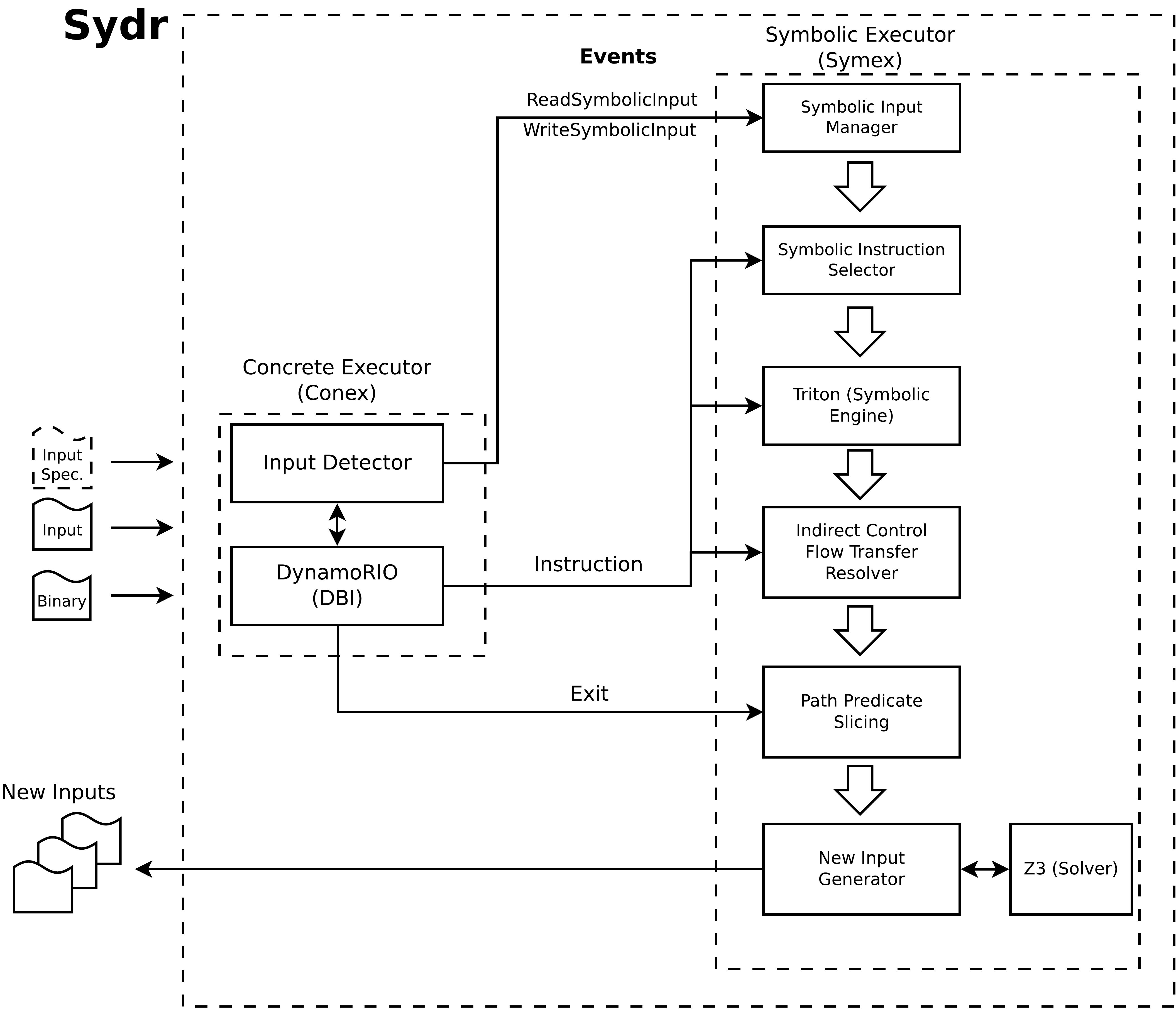}
    \caption{Sydr architecture.}
    \label{fig:sydr-arch}
\end{figure}

Fig.~\ref{fig:sydr-arch} presents Sydr architecture. Sydr performs symbolic
execution while program runs. Concrete and symbolic execution are separated
into two processes communicating via shared memory. \textit{Concrete Executor}
places \textit{events} in shared memory that are later processed by
\textit{Symbolic Executor}.

\textit{Concrete Executor (Conex)} has two components: \textit{Input Detector}
and \textit{DynamoRIO}. \textit{Input Detector} recognizes system calls and
library functions that handle input data specified by user. When such system
call is detected \textit{Conex} sends an event to \textit{Symbolic Executor
(Symex)}. This event (\texttt{ReadSymbolicInput}) holds information used by
\textit{Symex} to create new symbolic variables. \texttt{WriteSymbolicInput}
event allows to track symbolic variables when data
are stored to disk. DynamoRIO implements dynamic binary instrumentation.
\textit{Input Detector} also requires DBI to hook system and library calls. DBI
collects information about executed instruction: address, opcode, explicit and
implicit operands with their concrete values. This information is sent as
\texttt{Instruction} event to \textit{Symex} for symbolic execution.

\textit{Symbolic Executor (Symex)} handles events from \textit{Conex} to perform
symbolic execution. \textit{Symbolic Input Manager} is responsible for creating
symbolic variables. It updates symbolic registers, memory, and files states when
\texttt{Read/WriteSymbolicInput} events occur. Manager also contains concrete
values of input data corresponding to symbolic variables. These values are
needed while producing new inputs. When \textit{Conex} detects first read from
symbolic input, it starts passing \texttt{Instruction} events to \textit{Symex}.
These events firstly go through \textit{Symbolic Instruction Selector}. Selector
chooses just instructions having at least one symbolic operand (explicit or
implicit). These selected instructions are executed symbolically by Triton.
\textit{Indirect Control Flow Transfer Resolver} detects indirect control
transfer instructions, determines possible control flow target addresses, and
constructs path predicate constraints for them. \texttt{Exit} event
passed to \textit{Symex} signals that concrete execution is stopped. At this
point path constraint building is finished and \textit{Path Predicate Slicing}
component starts to perform algorithm described in \textbf{\nameref{Slicing}}
section. \textit{New Input Generator} component inverts branch conditions in
path predicate (including indirect control flow jumps/calls) to produce new
inputs. To invert each branch a corresponding SMT solver (we use
Z3~\cite{demoura08, z3}) query is
formed. In each query we only use those symbolic variables that affect a target
branch to be inverted, i.e. the other parts of input data stay unchanged. The
set of inputs from \textit{New Input Generator} is provided to user.

Sydr supports parallel inversion of branches. We build a complete path predicate
first and then solve SMT queries in parallel threads. Moreover, we terminate
each solver query by a specified timeout. We could have been inverting some
branches during the path predicate construction, but further research is needed.

\section{Evaluation}

\label{Evaluation}

We evaluated Sydr on a set of single-threaded 64-bit Linux executables~\cite{sydr-benchmark}.
We leave multi-threaded programs evaluation for future research.
For evaluation we used the server with the following specification:
processor AMD EPYC 7702 (128 cores), 256G~RAM.
We also checked the correctness of generated inputs.
If new input has the same execution trace as original except the last branch,
that should be in inverted direction, this input is correct.
We developed a tool based on DynamoRIO to verify the inputs correctness.
In tables below column named \textbf{Correct} represents the number of correct
inputs. The column named \textbf{SAT} shows the amount of satisfiable solver
\textbf{Queries}.
Each query is an attempt to invert branch (change control flow direction). The column named \textbf{Branches} is the number of symbolic branches in path predicate.
It should be noted that the total number of branches can be less than the number
of queries, because each jump table is considered as one branch and produces
multiple queries.
Sydr inverts branches from first to last in path predicate. Each test is
executed up to 2 hours. We limit path predicate construction time to 20
minutes.

\begin{table}[htbp]
\caption{Results With All Proposed Techniques}
\begin{center}
\scriptsize
\begin{tabular}{l >{\columncolor[gray]{0.9}}r r r r r}
\toprule
    \textbf{Application}&\textbf{Correct}&\textbf{SAT}&\textbf{Queries}&\textbf{Branches}&\textbf{Time} \\
    bzip2recover&2101&2101&5131&5131&51m3s \\
    cjpeg&50&50&197&8010&120m \\
    faad&426&430&652&458145&120m \\
    foo2lava&27&31&6127&910725&120m \\
    hdp&809&1037&3828&67476&120m \\
    jasper&6766&6798&18207&837669&120m \\
    libxml2&545&1069&17532&53699&120m \\
    minigzip&3896&7569&8977&8977&29m42s \\
    muraster&3227&3228&4726&7102&120m \\
    pk2bm&182&183&3673&3673&21m39s \\
    pnmhistmap\_pgm&17088&17089&25446&967187&120m \\
    pnmhistmap\_ppm&106&107&8247&8121&28m52s \\
    readelf&639&739&6141&64196&120m \\
    yices-smt2&2114&2699&9647&19543&120m \\
    yodl&180&313&5201&4831&34m59s \\
\bottomrule
\end{tabular}
\label{tbl:no-shuffle}
\end{center}
\end{table}

We present results with all proposed techniques in Table~\ref{tbl:no-shuffle}.
Then we disable some method in order to determine its influence.

\begin{table}[htbp]
\caption{Results Without Path Predicate Slicing}
\begin{center}
\scriptsize
\begin{tabular}{l >{\columncolor[gray]{0.9}}r r r r r}
\toprule
    \textbf{Application}&\textbf{Correct}&\textbf{SAT}&\textbf{Queries}&\textbf{Branches}&\textbf{Time} \\
    bzip2recover&2101&2101&5131&5131&52m42s \\
    cjpeg&50&50&198&8010&120m \\
    faad&386&389&585&470588&120m \\
    foo2lava&27&31&6252&910725&120m \\
    hdp&116&464&2427&67475&120m \\
    jasper&1&1987&5639&837669&120m \\
    libxml2&130&1043&13520&53700&120m \\
    minigzip&425&3961&4183&8977&120m \\
    muraster&3234&3235&4739&7102&120m \\
    pk2bm&181&183&3672&3673&21m48s \\
    pnmhistmap\_pgm&3158&3159&4681&967187&120m \\
    pnmhistmap\_ppm&106&107&8247&8121&40m15s \\
    readelf&135&218&2046&64196&120m \\
    yices-smt2&13&521&2135&19543&120m \\
    yodl&26&313&5201&4831&43m24s \\
\bottomrule
\end{tabular}
\label{tbl:no-slicing}
\end{center}
\end{table}

Table~\ref{tbl:no-slicing} contains results without path predicate slicing. Slicing significantly increases accuracy of generated inputs.
For some programs (\textit{jasper}, \textit{minigzip}, \textit{hpd},
\textit{pnmhistmap} with .pgm file, \textit{yices-smt2}, \textit{readelf}) the amount
of correct branches increased in several times with path predicate slicing.
Still, for \textit{bzip2recover} and \textit{cjpeg} result stays the same.

\begin{table}[htbp]
\caption{Results With All Proposed Techniques (Branches Are Randomly Chosen)}
\begin{center}
\scriptsize
\begin{tabular}{l >{\columncolor[gray]{0.9}}r r r r r}
\toprule
    \textbf{Application}&\textbf{Correct}&\textbf{SAT}&\textbf{Queries}&\textbf{Branches}&\textbf{Time} \\
    bzip2recover&2101&2101&5131&5131&51m31s \\
    cjpeg&6&6&425&8010&120m \\
    faad&18&71&1234&470588&120m \\
    foo2lava&1&1&19&910725&120m \\
    hdp&24&48&427&67476&120m \\
    jasper&34&34&127&837669&120m \\
    libxml2&26&62&2918&53699&120m \\
    minigzip&3896&7569&8977&8977&18m26s \\
    muraster&102&125&246&7102&120m \\
    pk2bm&182&183&3672&3673&21m47s \\
    pnmhistmap\_pgm&537&537&819&967187&120m \\
    pnmhistmap\_ppm&106&107&8247&8121&27m \\
    readelf&61&70&833&64196&120m \\
    yices-smt2&1193&1543&5811&19543&120m \\
    yodl&180&313&5201&4831&39m3s \\
\bottomrule
\end{tabular}
\label{tbl:master}
\end{center}
\end{table}

Table~\ref{tbl:master} presents results with all proposed techniques and randomly chosen branches.
The amount of correct inputs for tests fitting in 2 hour limit is the same.
For other tests the number of correct inputs decreased because the complexity of
solver queries for randomly chosen branches increased.

\begin{table}[htbp]
\caption{Parallel Solving (Correct Inputs)}
\begin{center}
\begin{tabular}{l r r r r}
\toprule
    \multirow{2}{*}{\textbf{Application}}&\multicolumn{4}{c}{\textbf{Number of Threads}} \\
    &\textbf{1}&\textbf{2}&\textbf{4}&\textbf{8} \\
    cjpeg&50&54&113&113 \\
    faad&426&507&582&2803 \\
    foo2lava&27&27&32&32 \\
    hdp&809&1052&1441&1813 \\
    jasper&6766&9746&11965&12269 \\
    libxml2&545&545&545&545 \\
    muraster&3227&3450&3896&3968 \\
    pnmhistmap\_pgm&17088&19861&24360&24115 \\
    readelf&639&1244&1652&2018\\
    yices-smt2&2114&3002&4147&4157 \\
\bottomrule
\end{tabular}
\label{tbl:parallel-correct}
\end{center}
\end{table}

\begin{table}[htbp]
\caption{Parallel Solving (Time)}
\begin{center}
\begin{tabular}{l r r r r}
\toprule
    \multirow{2}{*}{\textbf{Application}}&\multicolumn{4}{c}{\textbf{Number of Threads}} \\
    &\textbf{1}&\textbf{2}&\textbf{4}&\textbf{8} \\
    bzip2recover&51m3s&25m57s&13m35s&10m12 \\
    cjpeg&---&---&63m12s&24m8s \\
    minigzip&29m42s&17m18s&9m13s&6m49s \\
    pk2bm&21m39s&11m21s&5m47s&3m1s \\
    pnmhistmap\_ppm&28m52s&14m20s&7m34s&4m14s \\
    yodl&34m59s&16m54s&9m14s&5m23s\\
\bottomrule
\end{tabular}
\label{tbl:parallel-time}
\end{center}
\end{table}

We evaluated how parallel solving influences input generation. We ran
benchmark with all proposed techniques using 1, 2, 4, and 8 solving threads.
Table~\ref{tbl:parallel-correct} shows how  parallel solving increases the
amount of correct generated inputs. This table displays tests that do not fit in 2 hour limit and don't invert all branches in path predicate.
For \textit{libxml2} test the number of correct inputs stayed the same, but the number of queries increased from 17532 to 38092.
Degradation of results for \textit{pnmhistmap\_pgm} (8 threads worse than 4
threads) could be explained by exhaustion of all CPU cores. We ran several tests in parallel.
Tool for testing the correctness of inputs ran in parallel too. Table~\ref{tbl:parallel-time} represents how time needed for analysis decreased with parallel solving.
There are only tests that fit in 2 hour time limit and invert all branches in path predicate.
We can see that test named \textit{cjpeg} using 4 or more threads fits in 2 hour
time limit.

\begin{table}[htbp]
\caption{Path Predicate Construction Time}
\begin{center}
\scriptsize
\begin{tabular}{l r r r r r >{\columncolor[gray]{0.9}}r}
\toprule
    \multirow{2}{*}{\textbf{Application}}&\textbf{Input}&\textbf{Branch}&\textbf{App}&\multicolumn{3}{c}{\textbf{Path Predicate Time}} \\
    &\textbf{Size}&\textbf{Count}&\textbf{Time}&\textbf{Base}&\textbf{Skip}&\textbf{X} \\
    bzip2recover&147b&5131&0.0018s&9s&5s&1.8 \\
    cjpeg&12K&8010&0.0017s&39s&16s&2.4 \\
    faad&33K&470588&0.0082s&46m35s&18m7s&2.6 \\
    foo2lava&34K&910725&0.0045s&22m32s&18m42s&1.2 \\
    hdp&530K&67478&0.0021s&1m6s&41s&1.6 \\
    jasper&198K&837669&0.0037s&---&14m11s&--- \\
    libxml2&453b&53699&0.0024s&1m5s&34s&1.9 \\
    minigzip&19K&8977&0.0023s&2m44s&58s&2.8 \\
    muraster&887b&7102&0.0024s&7s&3s&2.3 \\
    pk2bm&1.7K&3673&0.0018s&4s&2s&2.0 \\
    pnmhistmap\_pgm&198K&967187&0.0038s&14m37s&7m55s&1.8 \\
    pnmhistmap\_ppm&12K&8121&0.0021s&29s&11s&2.6 \\
    readelf&8.3K&64196&0.0019s&1m19s&36s&2.2 \\
    yices-smt2&2K&19543&0.0029s&26s&14s&1.9 \\
    yodl&280b&4831&0.0017s&21s&6s&3.5 \\
\bottomrule
\end{tabular}
\label{tbl:path-predicate}
\end{center}
\end{table}

The Table~\ref{tbl:path-predicate} contains evaluation of path predicate construction time.
The column \textbf{App Time} represents the running time of the program without instrumentation.
The column \textbf{Base} shows running time without skipping non-symbolic instructions.
Running time with skipping non-symbolic instructions is presented in the column \textbf{Skip}.
Skipping non-symbolic instructions makes path predicate building 1.2--3.5 times faster.
Path predicate building for \textit{jasper} didn't complete for 24 hours. We
should investigate the reasons for that later.

\begin{table}[htbp]
\caption{Results Without Indirect Control Transfers Resolving}
\begin{center}
\scriptsize
\begin{tabular}{l >{\columncolor[gray]{0.9}}r r r r r}
\toprule
    \textbf{Application}&\textbf{Correct}&\textbf{SAT}&\textbf{Queries}&\textbf{Branches}&\textbf{Time} \\
    bzip2recover&2101&2101&5131&5131&51m8s \\
    cjpeg&50&50&197&7986&120m \\
    faad&427&431&653&422272&120m \\
    foo2lava&27&31&6119&910725&120m \\
    hdp&815&1050&3851&67383&120m \\
    jasper&6572&6604&17710&837670&120m \\
    libxml2&545&1085&16232&53548&120m \\
    minigzip&3896&7569&8977&8977&31m34s \\
    muraster&2652&3861&4998&6018&120m \\
    pk2bm&181&183&3673&3673&21m30s \\
    pnmhistmap\_pgm&17062&17063&25410&967187&120m \\
    pnmhistmap\_ppm&106&107&8058&8058&27m \\
    readelf&629&727&5815&64093&120m \\
    yices-smt2&2056&2596&9183&19386&120m \\
    yodl&159&275&4795&4795&33m47s \\
\bottomrule
\end{tabular}
\label{tbl:no-jump-tables}
\end{center}
\end{table}

Results of the tool application without indirect control flow transfers are shown in Table~\ref{tbl:no-jump-tables}.
Only a few tested programs have symbolic indirect jumps:
\textit{faad}, \textit{muraster}, \textit{readelf},
\textit{yices}, \textit{yodl}. Other programs can be used to evaluate how pure jump table detection
algorithm affects the analysis performance.

The programs with indirect jumps have more detected symbolic branches.
New inputs that lead to previously undiscovered paths were generated for these programs.
The largest number of indirect jumps was found on \textit{muraster}~-- near a 1000
symbolic branches. These queries were quite difficult for solver to process,
so there are less processed queries in the same time. On the contrary, the number of
incorrectly generated inputs (that do not actually invert the target branch) has
significantly decreased. Without processing indirect jumps, the program path predicate
is missing some of the branches that depend on the input. Inversion of a branch after such
missed jumps will not consider its condition, which in turn will lead to the generation
of the wrong input. For instance, we cannot correctly invert branch in a switch
case. Thus, resolving indirect control transfers allows to increase an
analysis accuracy.

The number of discovered branches and
processed queries for programs without symbolic indirect jumps did not change
significantly. Therefore the implemented jump table detection
mechanism does not reduce the tool performance.

\section{Future Work}

We plan to continue research in improving dynamic symbolic execution.
There are several interesting areas to research:
\begin{itemize}
    \item Modeling \textit{function semantics} in symbolic execution could
      increase accuracy and possibly speed up DSE
      (\texttt{tolower}/\texttt{toupper} are interesting because they constrain
      a symbol case).
    \item \textit{Symbolic memory model}~\cite{cha12} could provide new symbolic states interesting for futher analysis.
    \item Using \textit{Z3-solver tactics} could possibly decrease time spent in solver.
    \item Developing \textit{light-weight security predicates} to find some
        types of dangerous vulnerabilities.
\end{itemize}

We have already partially developed security predicates, which find
critical errors, such as null
pointer dereference and out of bounds access vulnerabilities.
In future we plan to research integer overflow, wraparound, and
some other dangerous types of critical defects.

Besides, during evaluation we independently found bugs in some programs~\cite{goblin, faad},
including commercial NTFS support module for UEFI. Furthermore, we plan to
improve bug detection in our analysis: iteratively launching application on
various inputs, testing generated inputs for hangs.

\section{Conclusion}

We have presented Sydr, a tool for dynamic symbolic execution that embodies
the best techniques to analyze real world programs. We designed it in a way to
provide an independence from restrictions imposed by instrumentation
platform and target programs. Our evaluation results showed that all considered
methods are crucial for accuracy and performance. The symbolic engine ASTs
simplification and skipping execution of non-symbolic instructions
enhance analysis efficiency. Path predicate slicing,
indirect control transfer resolving, and maintenance of thread-based symbolic
states allows us to significantly increase the analysis soundness and expand
the boundaries of our tool applicability.

\printbibliography

@article{baldoni18,
  author    = {Baldoni, Roberto and Coppa, Emilio and D'Elia, Daniele Cono and Demetrescu, Camil and Finocchi, Irene},
  title     = {A Survey of Symbolic Execution Techniques},
  journal   = {ACM Computing Surveys},
  volume    = {51},
  number = {3},
  articleno = {50},
  publisher = {ACM},
  doi = {10.1145/3182657},
  year = {2018}
}

@inproceedings{cha12,
 author = {Cha, Sang Kil and Avgerinos, Thanassis and Rebert, Alexandre and Brumley, David},
 title = {Unleashing {{Mayhem}} on Binary Code},
 booktitle = {Proceedings of the 2012 IEEE Symposium on Security and Privacy},
 series = {SP~'12},
 year = {2012},
 pages = {380--394},
 numpages = {15},
 doi = {10.1109/SP.2012.31},
 publisher = {IEEE Computer Society},
}

@inproceedings{godefroid08,
  title={Automated Whitebox Fuzz Testing},
  author={Godefroid, Patrice and Levin, Michael Y. and Molnar, David A.},
  booktitle={NDSS},
  volume={8},
  pages={151--166},
  year={2008},
  url={https://www.microsoft.com/en-us/research/publication/automated-whitebox-fuzz-testing/},
}

@article{king76,
 author = {King, James C.},
 title = {Symbolic Execution and Program Testing},
 journal = {Communications of the ACM},
 volume = {19},
 number = {7},
 year = {1976},
 pages = {385--394},
 numpages = {10},
 doi = {10.1145/360248.360252},
 publisher = {ACM}
}

@article{padaryan15,
  author={Padaryan, V.~A. and Kaushan, V.~V. and Fedotov, A.~N.},
  title={Automated exploit generation for stack buffer overflow vulnerabilities},
  journal={Programming and Computer Software},
  year={2015},
  volume={41},
  number={6},
  pages={373--380},
  doi={10.1134/S0361768815060055}
}

@inproceedings{schwartz10,
  title={All You Ever Wanted to Know about Dynamic Taint Analysis and Forward
         Symbolic Execution (but Might Have Been Afraid to Ask)},
  author={Schwartz, Edward J and Avgerinos, Thanassis and Brumley, David},
  booktitle={2010 IEEE Symposium on Security and Privacy},
  pages={317--331},
  year={2010},
  doi={10.1109/SP.2010.26}
}

@inproceedings{shoshitaishvili16,
 author = {Yan Shoshitaishvili and Ruoyu Wang and Christopher Salls and Nick Stephens and Mario Polino and Andrew Dutcher and John Grosen and Siji Feng and Christophe Hauser and Christopher Kruegel and Giovanni Vigna},
 booktitle = {2016 IEEE Symposium on Security and Privacy (SP)},
 title = {{{SOK}}: (State of) The Art of War: Offensive Techniques in Binary Analysis},
 year = {2016},
 pages = {138-157},
 doi = {10.1109/SP.2016.17},
}

@misc{smt-lib,
  author = {Clark Barrett and Pascal Fontaine and Cesare Tinelli},
  title = {{The SMT-LIB Standard: Version 2.6}},
  institution = {Department of Computer Science, The University of Iowa},
  year = 2017,
  url = {www.SMT-LIB.org},
}

@inproceedings{saudel15,
  author    = {Saudel, Florent and Salwan, Jonathan},
  title     = {{{Triton}}: A Dynamic Symbolic Execution Framework},
  booktitle = {Symposium sur la s{\'{e}}curit{\'{e}} des technologies de l'information
               et des communications},
  series    = {SSTIC},
  pages     = {31--54},
  year      = {2015},
  url       = {https://triton.quarkslab.com/files/sstic2015_slide_en_saudel_salwan.pdf}
}

@inproceedings{cadar06,
  author = {Cadar, Cristian and Ganesh, Vijay and Pawlowski, Peter M. and Dill, David L. and Engler, Dawson R.},
  title = {{{EXE}}: Automatically Generating Inputs of Death},
  year = {2006},
  publisher = {ACM},
  doi = {10.1145/1180405.1180445},
  booktitle = {Proceedings of the 13th ACM Conference on Computer and Communications Security},
  pages = {322–335},
  series = {CCS '06}
}

@inproceedings{cadar08,
  title={{{KLEE}}: Unassisted and Automatic Generation of High-Coverage Tests
         for Complex Systems Programs},
  author={Cadar, Cristian and Dunbar, Daniel and Engler, Dawson R},
  booktitle={OSDI},
  volume={8},
  pages={209--224},
  year={2008},
  url={https://static.usenix.org/events/osdi08/tech/full_papers/cadar/cadar.pdf}
}

@inproceedings{demoura08,
  author={de Moura, Leonardo and Bj{\o}rner, Nikolaj},
  title={{{Z3}}: An Efficient {{SMT}} Solver},
  booktitle={Tools and Algorithms for the Construction and Analysis of Systems},
  year={2008},
  publisher={Springer Berlin Heidelberg},
  doi = {10.1007/978-3-540-78800-3_24},
  pages={337--340},
}

@phdthesis{bruening04,
  title={Efficient, Transparent, and Comprehensive Runtime Code Manipulation},
  author={Bruening, Derek},
  year={2004},
  school={Massachusetts Institute of Technology, Department of Electrical
          Engineering and Computer Science},
  url = {https://www.burningcutlery.com/derek/docs/phd.pdf}
}

@techreport{molnar07,
  title={{{Catchconv}}: Symbolic execution and run-time type inference for integer conversion errors},
  author={Molnar, David A and Wagner, David},
  institution={UC Berkeley EECS},
  number={UCB/EECS-2007-23},
  year={2007},
  url={https://digitalassets.lib.berkeley.edu/techreports/ucb/text/EECS-2007-23.pdf}
}

@article{chipounov12,
  author = {Chipounov, Vitaly and Kuznetsov, Volodymyr and Candea, George},
  title = {The {{S2E}} Platform: Design, Implementation, and Applications},
  year = {2012},
  publisher = {ACM},
  volume = {30},
  number = {1},
  pages={1--49},
  doi = {10.1145/2110356.2110358},
  journal = {ACM Transactions on Computer Systems (TOCS)},
  articleno = {2},
}

@inproceedings{yun18,
  title={{{QSYM}}: A Practical Concolic Execution Engine Tailored for Hybrid Fuzzing},
  author={Yun, Insu and Lee, Sangho and Xu, Meng and Jang, Yeongjin and Kim, Taesoo},
  booktitle={27th USENIX Security Symposium},
  pages={745--761},
  year={2018},
  url={https://www.usenix.org/system/files/conference/usenixsecurity18/sec18-yun.pdf}
}

@article{cifuentes01,
  title = {Recovery of jump table case statements from binary code},
  journal = {Science of Computer Programming},
  volume = {40},
  number = {2},
  pages = {171--188},
  year = {2001},
  doi = {10.1016/S0167-6423(01)00014-4},
  author = {Cifuentes, Cristina and Van Emmerik, Mike},
}

@article{weiser84,
  author={Weiser, Mark},
  journal={IEEE Transactions on Software Engineering},
  doi={10.1109/TSE.1984.5010248},
  title={Program Slicing},
  year={1984},
  volume={SE-10},
  number={4},
  pages={352--357},
}

@inproceedings{kang11,
  title={{{DTA++}}: Dynamic Taint Analysis with Targeted Control-Flow Propagation},
  author={Kang, Min Gyung and McCamant, Stephen and Poosankam, Pongsin and Song, Dawn},
  booktitle={Proceedings of the Network and Distributed System Security Symposium},
  series = {NDSS '11},
  year={2011}
}

@inproceedings{guo16,
  author = {Guo, Shengjian and Kusano, Markus and Wang, Chao},
  title = {{{Conc-ISE}}: Incremental Symbolic Execution of Concurrent Software},
  year = {2016},
  publisher = {ACM},
  doi = {10.1145/2970276.2970332},
  booktitle = {Proceedings of the 31st IEEE/ACM International Conference on Automated Software Engineering},
  pages = {531–542},
  series = {ASE 2016}
}

@inproceedings{gerasimov17,
  title={Anxiety: a dynamic symbolic execution framework},
  author={Gerasimov, Alexander and Vartanov, Sergey and Ermakov, Mikhail and Kruglov, Leonid and Kutz, Daniil and Novikov, Alexander and Asryan, Seryozha},
  doi={10.1109/ISPRAS.2017.00010},
  booktitle={2017 Ivannikov ISPRAS Open Conference (ISPRAS)},
  pages={16--21},
  year={2017},
  organization={IEEE}
}

@inproceedings{aslanyan20,
  title={{{BinSide}} : Static Analysis Framework for Defects Detection in Binary
         Code},
  author={Aslanyan, Hayk and Arutunian, Mariam and Keropyan, Grigor and
          Kurmangaleev, Shamil and Vardanyan, Vahagn},
  doi={10.1109/IVMEM51402.2020.00007},
  booktitle={2020 Ivannikov Memorial Workshop (IVMEM)},
  pages={9--14},
  year={2020},
  organization={IEEE}
}

@article{ivannikov14,
  title={Static analyzer {{Svace}} for finding defects in a source program code},
  author={Ivannikov, V.~P. and Belevantsev, A.~A. and Borodin, A.~E. and
          Ignatiev, V.~N. and Zhurikhin, D.~M. and Avetisyan, A.~I.},
  journal={Programming and Computer Software},
  volume={40},
  number={5},
  pages={265--275},
  year={2014},
  doi={10.1134/S0361768814050041}
}

@article{isaev10,
  title={The use of dynamic analysis for generation of input data that demonstrates critical bugs and vulnerabilities in programs},
  author={Isaev, I.~K. and Sidorov, D.~V.},
  journal={Programming and Computer Software},
  volume={36},
  number={4},
  pages={225--236},
  year={2010},
  doi={10.1134/S0361768810040055}
}

@inproceedings{fioraldi20,
  title={{{AFL++}}: Combining Incremental Steps of Fuzzing Research},
  author={Fioraldi, Andrea and Maier, Dominik and Ei{\ss}feldt, Heiko and Heuse, Marc},
  booktitle={14th USENIX Workshop on Offensive Technologies (WOOT 20)},
  year={2020},
  url={https://www.usenix.org/system/files/woot20-paper-fioraldi.pdf}
}

@inproceedings{balakrishnan05,
  author={Balakrishnan, Gogul and Gruian, Radu and Reps, Thomas and Teitelbaum, Tim},
  title={{{CodeSurfer/x86}}---A Platform for Analyzing x86 Executables},
  booktitle={Compiler Construction},
  year={2005},
  publisher={Springer Berlin Heidelberg},
  doi={10.1007/978-3-540-31985-6_19},
  pages={250--254},
}

@inproceedings{sargsyan19,
  author={Sargsyan, Sevak and Hakobyan, Jivan and Mehrabyan, Matevos and Mishechkin, Maxim and Akozin, Vitaliy and Kurmangaleev, Shamil},
  booktitle={2019 Ivannikov Memorial Workshop (IVMEM)},
  title={{{ISP-Fuzzer}}: Extendable Fuzzing Framework},
  year={2019},
  pages={68--71},
  organization={IEEE},
  doi={10.1109/IVMEM.2019.00017}
}

@book{howard06,
  title={The security development lifecycle},
  author={Howard, Michael and Lipner, Steve},
  volume={8},
  year={2006},
  publisher={Microsoft Press Redmond},
  url={http://msdn.microsoft.com/en-us/library/ms995349.aspx}
}

@standard{iso08,
  title={{{ISO/IEC}} 15408-3:2008: Information technology~-- Security techniques~--
         Evaluation criteria for IT security~-- Part 3: Security assurance
         components},
  year={2008},
  organization={ISO Geneva, Switzerland},
  url={https://www.iso.org/standard/46413.html}
}

@standard{gost16,
  title={{{GOST R}} 56939-2016: Information protection. Secure software development.
         General requirements},
  organization={National Standard of Russian Federation},
  year={2016},
  url={http://protect.gost.ru/document.aspx?control=7&id=203548},
}

@article{bessey10,
  author = {Bessey, Al and Block, Ken and Chelf, Ben and Chou, Andy and Fulton, Bryan and Hallem, Seth and Henri-Gros, Charles and Kamsky, Asya and McPeak, Scott and Engler, Dawson},
  title = {A Few Billion Lines of Code Later: Using Static Analysis to Find Bugs in the Real World},
  year = {2010},
  publisher = {ACM},
  volume = {53},
  number = {2},
  doi = {10.1145/1646353.1646374},
  journal = {Communications of the ACM},
  pages = {66–75},
}

@inproceedings{stephens16,
  title={Driller: Augmenting Fuzzing Through Selective Symbolic Execution},
  author={Stephens, Nick and Grosen, John and Salls, Christopher and Dutcher, Andrew and Wang, Ruoyu and Corbetta, Jacopo and Shoshitaishvili, Yan and Kruegel, Christopher and Vigna, Giovanni},
  booktitle={NDSS},
  volume={16},
  number={2016},
  pages={1--16},
  year={2016}
}

@misc{sydr-benchmark,
  title = {Sydr benchmark},
  url = {https://github.com/ispras/sydr-benchmark}
}

@misc{goblin,
  title = {Goblin bug},
  url = {https://github.com/m4b/goblin/issues/108}
}

@misc{faad,
  title = {Faad2 bug},
  url = {https://github.com/knik0/faad2/pull/65}
}

@misc{drpthread,
  author = {Bruening, Derek},
  title = {Issue for {{DynamoRIO}} libpthread support},
  url = {https://github.com/DynamoRIO/dynamorio/issues/2848}
}

@misc{drheap,
  author = {Bruening, Derek},
  title = {Issue for {{DynamoRIO}} heap slowdowns},
  url = {https://github.com/DynamoRIO/dynamorio/issues/2115}
}

@misc{dynamorio,
  author = {Bruening, Derek},
  title = {{{DynamoRIO}}: Dynamic Instrumentation Tool Platform},
  url = {https://github.com/DynamoRIO/dynamorio}
}

@misc{triton,
  author = {Salwan, Jonathan},
  title = {{{Triton}}: Dynamic Binary Analysis framework},
  url = {https://github.com/JonathanSalwan/Triton}
}

@misc{z3,
  author = {De Moura, Leonardo and Bj{\o}rner, Nikolaj},
  title = {The {{Z3}} Theorem Prover},
  url = {https://github.com/Z3Prover/z3}
}

@misc{claripy,
  title = {Claripy: An abstraction layer for constraint solvers},
  url = {https://github.com/angr/claripy}
}

\end{document}